\documentclass{article}
\usepackage{ijcai25}

\usepackage{times}
\usepackage{soul}
\usepackage{url}
\usepackage[hidelinks]{hyperref}
\usepackage[utf8]{inputenc}
\usepackage[small]{caption}
\usepackage{graphicx}
\usepackage{amsmath}
\usepackage{amsthm}
\usepackage{booktabs}
\usepackage{algorithm}
\usepackage{algorithmic}
\usepackage[switch]{lineno}
\usepackage{adjustbox}
\usepackage{tabularx}
\usepackage{makecell}
\usepackage{multirow}
\usepackage{amssymb}
\usepackage{svg}

\title{WDMIR: Wavelet-Driven Multimodal Intent Recognition}

\author{
Weiyin Gong$^{1,2}$
\and
Kai Zhang$^{1,}$\thanks{Corresponding author.}\and
Yanghai Zhang$^{1}$\and
Qi Liu$^{1}$\and
Xinjie Sun$^{1,2}$\and 
Junyu Lu$^{3}$\And \\
Linbo Zhu$^{3}$\\
\affiliations
$^1$State Key Laboratory of Cognitive Intelligence, University of Science and Technology of China\\
$^2$School of Computer Science, Liupanshui Normal University\\
$^3$Institute of Artificial Intelligence, Hefei Comprehensive National Science Center\\
\emails
\{weiyingong,yhzhang0612, xinjiesun,lujunyu\}@mail.ustc.edu.cn,
\{kkzhang08,qiliuql\}@ustc.edu.cn,
lbzhu@iai.ustc.edu.cn
}
 
\begin{document}

\maketitle
\begin{abstract}
    Multimodal intent recognition (MIR) seeks to accurately interpret user intentions by integrating verbal and non-verbal information across video, audio and text modalities. While existing approaches prioritize text analysis, they often overlook the rich semantic content embedded in non-verbal cues. This paper presents a novel \emph{Wavelet-Driven Multimodal Intent Recognition} \textbf{\emph{(WDMIR)}} framework that enhances intent understanding through frequency-domain analysis of non-verbal information. To be more specific, we propose: (1) a wavelet-driven fusion module that performs synchronized decomposition and integration of video-audio features in the frequency domain, enabling fine-grained analysis of temporal dynamics; (2) a cross-modal interaction mechanism that facilitates progressive feature enhancement from bimodal to trimodal integration, effectively bridging the semantic gap between verbal and non-verbal information. Extensive experiments on MIntRec demonstrate that our approach achieves state-of-the-art performance, surpassing previous methods by 1.13\% on accuracy. Ablation studies further verify that the wavelet-driven fusion module significantly improves the extraction of semantic information from non-verbal sources, with a 0.41\% increase in recognition accuracy when analyzing subtle emotional cues.
\end{abstract}
\section{Introduction}
Intent recognition is a key aspect of human-computer interaction, and its core goal is to enable machines to accurately grasp user intent and thus provide users with better service~\cite{zhang2019interactive,huang2023effective,qiu2024application}. Recently, multimodal intent recognition has been used to understand the user's intent in more complex scenarios. Compared with unimodal intent, multimodal information fusion can improve intent recognition accuracy by making joint decisions ~\cite{soleymani2017multimodal,chen2021multimodal,zou2022divide,zhang2024leveraging,sun2025daskt}. 
\begin{figure}
  \centering
  \vspace{0.35cm}
  \includegraphics[width=0.98\linewidth]{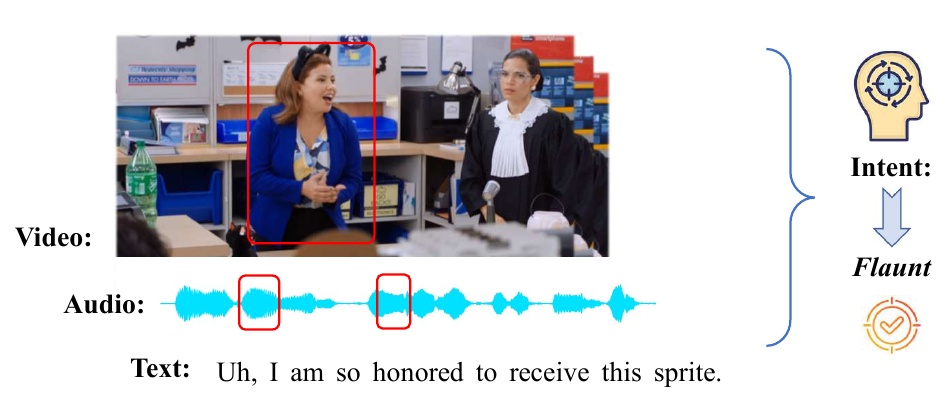}
  \vspace{-0.15cm}
  \caption{In the task of multimodal intent recognition, videos and audio contain relatively few key pieces of information, necessitating the in-depth mining of complementary information.}
  \vspace{-0.2cm}
  \label{figure 1}
\end{figure}

Researchers are currently paying greater attention to multimodal intent detection research as it focuses more on intricate real-world situations. To this end,~\cite{zhang2022mintrec} introduces the first multimodal intention recognition baseline dataset, MIntRec, and conducts experiments on models such as MulT~\cite{tsai2019multimodal}, MISA~\cite{hazarika2020misa}, and MAG-BERT~\cite{rahman2020integrating} to establish a baseline for intent recognition for subsequent research. Subsequently, in order to effectively integrate information from different modalities such as text, video, and audio, researchers have designed various models, all of which have achieved certain results in the field of multimodal intent recognition~\cite{zhang2022incorporating,sun2024contextual,huang2024sdif,zhou2024token}. Although existing methods have made some progress in multimodal intent recognition, the accuracy of multimodal analysis of user's intent is still limited due to the fact that the potential correlation between different modalities has not yet been fully explored, as well as the insufficiency of the video and audio modalities in semantic feature extraction. As shown in Figure \ref{figure 1}, when the speaker says, “Uh, I am so honored to receive this sprite,” we might initially infer that the speaker is joking. However, by carefully observing the speaker's facial expressions and analyzing their tone, we can discern that the speaker's true intention is to flaunt. so much so that multimodal intention recognition currently presents the following two challenges: first, how to deeply mine the semantic information in video and audio modalities; and second, how to efficiently align and fuse the features of text, video and audio modalities.

To address the first challenge, we propose a wavelet-driven approach. As far as we know, we are the first to introduce wavelet transform to drive the fusion of video and audio data information. Wavelet transform decomposes signals into low-frequency and high-frequency components. The low-frequency component captures the global characteristics of the signal, reflecting smooth trends and large-scale variations~\cite{satirapod2001approach}, while the high-frequency component focuses on local details, such as abrupt changes, fine structures, and rapid variations~\cite{lahmiri2014wavelet,li2023learning}. 

To address the second challenge, we designed collaborative representations and progressive fusion modules. These modules aim to enhance the alignment and integration between wavelet-driven non-verbal modalities and text modalities through cross-modal mechanisms, achieving a transition from bimodal to trimodal collaborative representations. Subsequently, the progressive fusion module is utilized for deep representation, leveraging complementary information at different stages to improve the accuracy of multimodal intent recognition. Our primary contributions in this study are:

\begin{itemize}
    \item We design wavelet-driven multimodal intent recognition methods to achieve fusion of nonverbal modal features in the frequency domain to improve the model's ability to understand and recognize multimodal data.
    \item We achieve cross-modal collaborative alignment to integrate multimodal information, and through progressive fusion, mine complementary information to enhance recognition accuracy.
    \item Our experiments significantly improved each metric on MIntRec and MELD-DA, validating the validity and generalizability of the method.
\end{itemize}
\section{Related Works}
\subsection{Multimodal Intent Recognition}
Multimodal intent recognition aims to extract the user intent from multimodal information. However, challenges remain when dealing with scenarios involving text, visual, and acoustic modalities.~\cite{zhang2022mintrec} introduced the multimodal intent recognition task and released a MIntRec dataset, which integrates textual, visual, and acoustic modalities to recognize user intent comprehensively.~\cite{zhou2024token} proposed a token-level contrastive learning method with modality-aware prompts that effectively integrates text, audio, and video modality features through similarity-based modality alignment and cross-modal attention.~\cite{huang2024sdif} introduced a shallow-to-deep interactive framework with data augmentation capabilities to address the modality alignment issue by gradually fusing multimodal features and incorporating ChatGPT-based data augmentation methods.~\cite{sun2024contextual} proposed a context-enhanced global contrastive method that alleviates biases and inconsistencies in multimodal intent recognition by using within-video and cross-video interactions and retrieval, combined with global context-guided contrastive learning.
\subsection{Multimodal Fusion Methods}
Multimodal fusion aims to integrate information from different modalities (e.g., text, video, and audio) to better understand and process information from multiple sources, thereby improving system performance~\cite{zhang2022graph}. For example, ~\cite{zadeh2017tensor} proposes a tensor fusion network that learns intra- and inter-modal dynamic features in an end-to-end manner.~\cite{tsai2019multimodal} addresses the problem of aligning multimodal sequences in different time steps by using directed pairwise cross-modal attention.~\cite{rahman2020integrating} Introduced multimodal adaptation gates to fine-tune BERT to address non-verbal modal input.~\cite{hazarika2020misa} maps each modality to two different subspaces to learn common and modality-specific features.
\subsection{Wavelet Transform}
Wavelet transform is a commonly used time-frequency analysis technique in signal processing that can convert information from the time domain to the frequency domain. The decomposed information exhibits both global and local characteristics. In recent years, many researchers have applied wavelet transforms to various fields.~\cite{li2023learning} introduced a wavelet fusion module for facial super-resolution tasks, addressing the issue in existing Transformer-based methods where global information integration often neglects relevant details, leading to blurring and limiting high-frequency detail recovery.~\cite{phutke2023blind} proposed an end-to-end blind image inpainting architecture with a wavelet query multi-head attention transformer block, effectively repairing images by using the wavelet coefficients processed to provide encoder features as queries, bypassing the damaged region prediction step.~\cite{sabry2024lung} applies wavelet transform to lung sound signal analysis, addressing the issues of noise contamination and artifact removal in lung sound signals. By performing a multi-scale analysis of the lung sound signals, effective features are extracted, improving the accuracy of lung disease classification.~\cite{frusque2024robust} introduced wavelet techniques for denoising audio information. Our method mainly uses wavelet transform to decompose video and audio information at multiple levels, to realize the fusion of video and audio features in the frequency domain, and to realize the analysis of non-verbal modal information in the frequency domain to enhance its representation.
\section{Method} 
\begin{figure*}
  \centering
  \includegraphics[width=\linewidth]{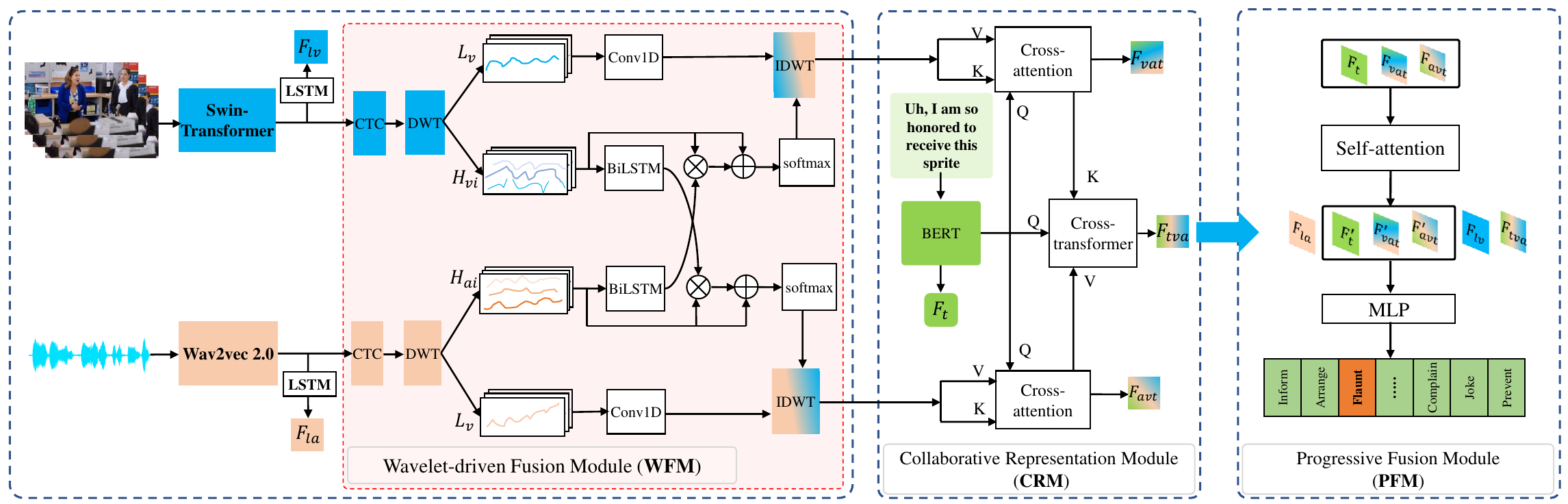}
  \caption{
 The overall framework of WDMIR primaril consists of the Wavelet-driven Fusion Module (WFM), Collaborative Representation Module (CRM), and Progressive Fusion Module (PFM), which are responsible for video and audio feature fusion, cross-modal collaborative representation, and enhancing intent recognition accuracy using multi-level features, respectively.}
 \label{figure 2}
\end{figure*}
\subsection{Task Description}
Let text, video, and audio data from three modalities are taken as inputs, their features are fused to predict user intent. Specifically, let the feature of the text modality be \( t \), the feature of the video modality be \( v \), and the feature of the audio modality be \( a \). By designing a multimodal fusion module \( \mathcal{F} \), these features are mapped into a unified representation.
\begin{equation}
     \mathbf{H} = \mathcal{F}(t, v, a) 
\end{equation}

Subsequently, a classifier \( \mathcal{F}_{\text{intent}} \) maps the fused features to the intent category space, generating a predicted distribution
\begin{equation}
    \hat{y} = \mathcal{F}_{\text{intent}}(\mathbf{H})
\end{equation}
where \( t \in \mathbb{R}^{d_t} \), \( v \in \mathbb{R}^{d_v} \), and \( a \in \mathbb{R}^{d_a} \) represent the features of the text, video, and audio modalities, respectively, with \( d_t \), \( d_v \), and \( d_a \) denoting the dimensions of each modality's features. \( \mathbf{H} \in \mathbb{R}^{d_h} \) is the fused feature representation; \( d_h \) is the dimension of the fused features; \( \hat{y} \in \mathbb{R}^{c} \) represents the predicted probability distribution over \( c \) intent categories.
\subsection{Framework Overview}
The overall framework of WDMIR is shown in Figure \ref{figure 2}. It mainly consists of Wavelet-driven Fusion Module (WFM), Collaborative Representation Module (CRM) and Progressive Fusion Module (PFM). WFM can realize the feature fusion of video and audio in frequency domain by multilevel decomposition of video and audio in the frequency domain. CRM utilizes the cross-modal mechanism to realize the collaborative representation of text modality and wavelet-driven video and audio modality to realize the collaborative representation of two modalities to three modalities. PFM improves the accuracy of intent recognition using multilevel feature progressive fusion representation.
\subsection{Feature Extraction}
For text modality, in order to fully utilize the semantic information in the text~\cite{liu2023enhancing}, we use pre-trained model BERT~\cite{lee2018pre} to encode the text and extract the last hidden layer as the feature representation of the text.
\begin{equation}
    F_t = \text{BERTEmbeding}(t)
\end{equation}
where $t$ is the input conversation text, $F_t$ is the text feature extracted through BERTEmbedding.

For the video modality, we follow the research approach in~\cite{zhou2024token} and choose the Swin-Transformer~\cite{liu2021swin}, which performs excellently in the field of computer vision, to sample the video frame by frame and perform pre-training, extracting the last hidden layer to represent the visual semantic information.
\begin{equation}
    F_v = \text{Swin-Transformer}([f_1,f_2,..,f_{lv}])
\end{equation}
where $f_i$ is the $i$-th frame in each video segment, $lv$ is the number of frames sampled from the video, and $F_v$ is the visual semantic information extracted by Swin-Transformer.

For audio modality, we adopt the method~\cite{zhang2022mintrec} to pre-train the model using Wav2Vec 2.0~\cite{baevski2020wav2vec} to extract the output of the last hidden layer as the audio feature representation.
\begin{equation}
    F_a = \text{Wav2Vec 2.0}(a)
\end{equation}
where $a$ is the audio sequence corresponding to the video clip and $F_a$ is the audio feature extracted by Wav2Vec 2.0.
\subsection{Wavelet-driven Fusion Module (WFM)}
WFM is able to decompose audio and video signals over sequences into the frequency domain, where fine-grained feature fusion of non-verbal modalities is achieved.Before further analyzing the video and audio features, we perform sequence alignment of the video and audio features using the CTC model~\cite{graves2006connectionist}.
\begin{equation}
    {V, A} = CTC({F_v,F_a})
\end{equation}
where $V$ and $A$ represent the video and audio features after sequence alignment, respectively.

In order to facilitate the fusion of non-verbal modal information in the frequency domain. We choose Haar wavelet bases to perform 3-level 1-dimensional DWT transform on audio and video aligned in sequences to obtain the high-frequency and low-frequency components of video and audio information, which correspond to local and global features, respectively.
\begin{equation}
\begin{aligned}
    (L_a,H_{ai}) = DWT(A)\\
    (L_v,H_{vi}) = DWT(V)
\end{aligned}
\end{equation}
Where $L_a$ and $L_v$ are the low-frequency components of the audio and video features obtained through multilevel wavelet decomposition, respectively, and $H_{ai}$ and $H_{vi}$ are the $i$-th high-frequency components of the audio and video features after multilevel decomposition, respectively.

To better extract global information from video and audio data, we employ a one-dimensional convolution with a kernel size of 3 to strengthen the representation of low-frequency components, further enhancing the model's understanding of global information.
\begin{equation}
\begin{aligned}
    L^{'}_a=Conv1D(L_a)\\
    L^{'}_v=Conv1D(L_v)
\end{aligned}
\end{equation}
where $L^{'}_a$ and $L^{'}_v$ are the low-frequency components of audio and video enhanced by one-dimensional convolution.

For the multilevel high-frequency components obtained from wavelet decomposition, we concatenate the components at each level sequentially to construct the total high-frequency components for each modality. This approach integrates high-frequency information from different levels, enhancing the model's ability to capture fine-grained features.
\begin{equation}
    \begin{aligned}
        F_{hv} = cat(H_{v1},H_{v2},H_{v3})\\
        F_{ha} = cat(H_{a1},H_{a2},H_{a3})
    \end{aligned}
\end{equation}
where $F_{hv}$ and $F_{ha}$ are the total high-frequency components of video and audio, respectively. 

To better facilitate the fusion of video and audio information, we use the BiLSTM model to map audio to the video space and video to the audio space, respectively.
\begin{equation}
\begin{aligned}
    H^{'}_v=BiLSTM(F_{hv})\\
    H^{'}_a=BiLSTM(F_{ha})
\end{aligned}
\end{equation}
where $H^{'}_v$ is the visual feature obtained from the high-frequency component of the video through BiLSTM; $H^{'}_a$ is the audio feature obtained from the high-frequency component of the audio through BiLSTM.\\
To better leverage the complementary information between modalities and enhance feature representation, we designed a frequency domain interaction module.
\begin{equation}
    \begin{aligned}
       H_{AV}=softmax(F_{ha} \odot H^{'}_v + F_{ha})\\
       H_{VA}=softmax(F_{hv} \odot H^{'}_a + F_{hv})
    \end{aligned}
\end{equation}
where $H_{AV}$ is the audio feature obtained from the interaction between audio and video, $H_{VA}$ is the video feature obtained from the interaction between video and audio, and $\odot$ denotes the element-wise multiplication.

To better integrate the interaction-enhanced features with their respective low-frequency features, we perform an inverse wavelet transform to reconstruct the audio and video modalities, resulting in fused audio and video features.
\begin{equation}
    \begin{aligned}
       F_{AV}=IDWT(cat(L^{'}_a,H_{AV}))\\
       F_{VA}=IDWT(cat(L^{'}_v,H_{VA}))
    \end{aligned}
\end{equation}
where $F_{AV}$ is the reconstructed and enhanced audio modality feature, and $F_{VA}$ is the reconstructed video modality feature.
\subsection{Collaborative Representation Module (CRM)}
The text modality serves as the primary source of information for intent recognition. We treat the text modality as the main modality and perform pairwise interactions with the enhanced audio and video modalities separately. Cross-modal attention mechanisms are employed to achieve alignment and feature fusion. Subsequently, the audio and video modalities, weighted by the text features, are further interacted with the text modality, enabling deep fusion of the three modalities.
\begin{table*}
    \centering
    \renewcommand{\arraystretch}{1.2}
    \begin{adjustbox}{max width=\textwidth}
    \begin{tabularx}{\textwidth}{lXXXXXXXX}
        \toprule
        \multirow{2}{*}{\textbf{Methods}} & \multicolumn{4}{c}{\textbf{MIntRec}} & \multicolumn{4}{c}{\textbf{MELD-DA}} \\
        \cmidrule(lr){2-5} \cmidrule(lr){6-9}
        & \textbf{ACC} & \textbf{WF1} & \textbf{WP} & \textbf{R}
        & \textbf{ACC} & \textbf{WF1} & \textbf{WP} & \textbf{R} \\
        \midrule
        MAG-BERT      & 72.65 & 72.16 & 72.53 & 69.28 & 60.63 & 59.36 & 59.80 & 50.01 \\
        MISA          & 72.29 & 72.38 & 73.48 & 69.24 & 59.98 & 58.52 & 59.28 & 48.75  \\
        MulT          & 72.52 & 72.31 & 72.85 & 69.24 & 60.36 & 59.01 & 59.44 & 49.93 \\
        TCL-MAP       & 73.62 & 73.31 & 73.72 & 70.50 & \underline{61.75} & \underline{59.77} & \underline{60.33} & \underline{50.14}  \\
        SDIF-DA$*$      & \underline{73.93} & \underline{73.89} & \underline{74.18} & \underline{71.66} & $--$   & $--$   & $--$ & $--$  \\
        CAGC          & 73.39 & $--$ & $--$  & 70.39  & $--$   & $--$   & $--$ & $--$  \\
         \midrule
        WDMIR (Our)            & \textbf{75.06}      &\textbf{74.96}       &\textbf{75.26}        &\textbf{72.65}       &\textbf{62.16} & \textbf{60.94} &\textbf{61.44}  &\textbf{51.36}\\
        $\triangle$ &1.13$\uparrow$  &1.07$\uparrow$  &1.08$\uparrow$ &0.99$\uparrow$  &0.41$\uparrow$  &1.17$\uparrow$  &1.11$\uparrow$ &1.22$\uparrow$  \\
        \bottomrule
    \end{tabularx}   
    \end{adjustbox}
    \caption{The experimental results of our method on the MIntRec and MELD-DA datasets are as follows. $\triangle$ represent the comparison results between our method and the previous best method, bold indicates the best results, underline denotes the second best results, $\uparrow$ denotes the improvement effect, asterisks * indicate the results from our re-experimentation, and all other results are sourced from published papers.}
    \label{table 1}
\end{table*}

Considering that the text modality contains the primary information, we perform sequence alignment on the fused video and audio features through a cross-attention mechanism~\cite{huang2024sdif}. We choose the text modality $F_t$ as the query $Q$, and the fused video $F_{VA}$ and audio modality $F_{AV}$ as the key and value $K$ and $V$.
\begin{equation}
    \begin{aligned}
       Cross-attention(Q,K,V)=softmax(\frac{QK^{T}}{\sqrt{d_k}})V\\
       F_{vat} = Cross-attention(F_t,F_{VA},F_{VA})\\
       F_{avt} = Cross-attention(F_t,F_{AV},F_{AV})
    \end{aligned}
\end{equation}
where $F_{vat}$ and $F_{avt}$ denote the weighted video and audio features, respectively.

Then $F_t$ as the query vector Q, $F_{vat}$ as the key vector K, and $F_{avt}$ as the value vector V are used to realize the trimodal co-representation via cross-transformer~\cite{tsai2019multimodal}.
\begin{equation}
    F_{tva} = softmax(\frac{{F_t}{F_{vat}}^{T}}{\sqrt{d_k}}){F_{avt}}
\end{equation}
where $F_{tva}$ denotes the three modal co-representation features of text, video, and audio. 
\subsection{Progressive Fusion Module (PFM)}
PFM improves the accuracy of multimodal intent recognition mainly through multilevel feature progressive fusion.To focus on the fused text, video, and audio model features from different visions. We stack $F_t$, $F_{vat}$ and $F_{avt}$ to obtain the matrix $F_m$.
\begin{equation}
    F_m = [F_t, F_{vat},F_{avt}]
\end{equation}

Then, the stacked $F_m$ is enhanced through self-attention~\cite{vaswani2017attention,zhang2021eatn} for feature enhancement.
\begin{equation}
    F = Self-attention(F_m)
\end{equation}
where $F=[F^{'}_{t},F^{'}_{vat}, F^{'}_{avt}]$ denotes the enhanced features.

We utilize LSTM to learn the encoded video and audio information, taking the hidden state of the last layer as output to prevent the loss of key features during the processing of video and audio characteristics, thereby enhancing the model's understanding of complex data.
\begin{equation}
    \begin{aligned}
        F_{lv} = LSTM(F_v)\\
        F_{la} = LSTM(F_a)
    \end{aligned}
\end{equation}
where $F_{lv}$ and $F_{la}$ denote the video features and audio features obtained by LSTM, respectively.

We concatenate the features obtained from multiple layers and map them to the output space through an MLP to get the final output $\hat{y}$.
\begin{equation}
    \hat{y} = MLP(cat(F_{lv},F_{la},F^{'}_{t},F^{'}_{vat}, F^{'}_{avt},F_{tva}))
\end{equation}

To make the output better approximate the true distribution, we use a cross-entropy loss
function to measure the difference between the predicted and true values.
\begin{equation}
    \mathcal{L} = -\frac{1}{N} \sum_{i=1}^{N} \sum_{j=1}^{C} y_{ij} \log(\hat{y}_{ij})
\end{equation}
where $\mathcal{L}$ is the total loss function of the model. $N$ is the number of samples and $C$ is the number of intended categories. $y_{ij}$ is the true label of the $i$-th sample in the $j$th category. $\hat{y}_{ij}$ is the probability distribution of the $i$-th sample belonging to the $j$-th category as predicted by the model.
\section{Experiments}
\subsection{Datasets}
We conduct experiments on two datasets, MIntRec~\cite{zhang2022mintrec} and MELD-DA~\cite{saha2020towards}.
MIntRec is a multimodal intent dataset containing text, video, and audio, with 2224 samples and 20 intent categories. It includes 1334, 445, and 445 samples for training, validation, and testing, respectively.
MELD-DA is a multi-round emotion conversation dataset containing text, video, and audio, with 9988 samples and 12 emotion conversation behavior labels. It includes 6991, 999, and 1998 samples for training, validation, and testing, respectively.
\subsection{Baselines} \label{Baselines}
In our experiments, we will use state-of-the-art multimodal fusion methods as baselines: (1) MISA~\cite{hazarika2020misa} projects each modality into two different subspaces to learn the fusion of common features and unique attributes of different modalities. (2) MulT~\cite{tsai2019multimodal} potentially converts one modality to another for feature fusion by directing paired cross-modal attention. (3) MAG-BERT~\cite{rahman2020integrating} introduces a multimodal adaptation gate as an accessory to fine-tune BERT to enable it to handle non-verbal data for multimodal data fusion. (4) TCL-MAP~\cite{zhou2024token} uses the similarity of modalities to design a multimodal perceptual cueing module for modal alignment, and uses a cross-modal attention mechanism to generate modal perceptual cues for multimodal fusion. (5) SDIF-DA~\cite{huang2024sdif} enhances text data through ChatGPT, then designs interaction modules from shallow to deep and gradually aligns and fuses features between different modalities effectively. (6) CAGC~\cite{sun2024contextual} enhances the capture of global contextual features by mining the contextual interaction information within and between videos, thus effectively solving the problems of perceptual bias and inconsistency of multimodal representations.
\subsection{Evaluation Metrics} Based on previous work ~\cite{zhou2024token}, we use Accuracy (ACC), Weighted F1 Score (WF1), Weighted Precision (WP), and Recall (R) as the evaluation metrics of the model. The impact of sample unevenness on model performance is reduced by a weighted average of the number of samples in each category by WF1 and WP.
\subsection{Experimental Settings} In our experiments, bert-base-uncased\footnote{\url{https://huggingface.co/google-bert/bert-base-uncased}} and wav2vec2-base-960h\footnote{\url{https://huggingface.co/facebook/wav2vec2-base-960h}} from the Huggingface are used as the pre-training models for extracting mentioned text and audio features. Video features are extracted from the Torchvision library by using swin\_b pre-trained on ImageNet1K. Adam~\cite{loshchilov2017decoupled} as an optimization parameter throughout the experiment. The training batch size is 16, and the validation and test batch sizes are both 8.Our code is available\footnote{\url{https://github.com/gongweiyin/WDMIR}}.
\begin{figure*}
    \centering
    \begin{minipage}[t]{0.48\textwidth}
        \centering
        \includegraphics[width=\linewidth]{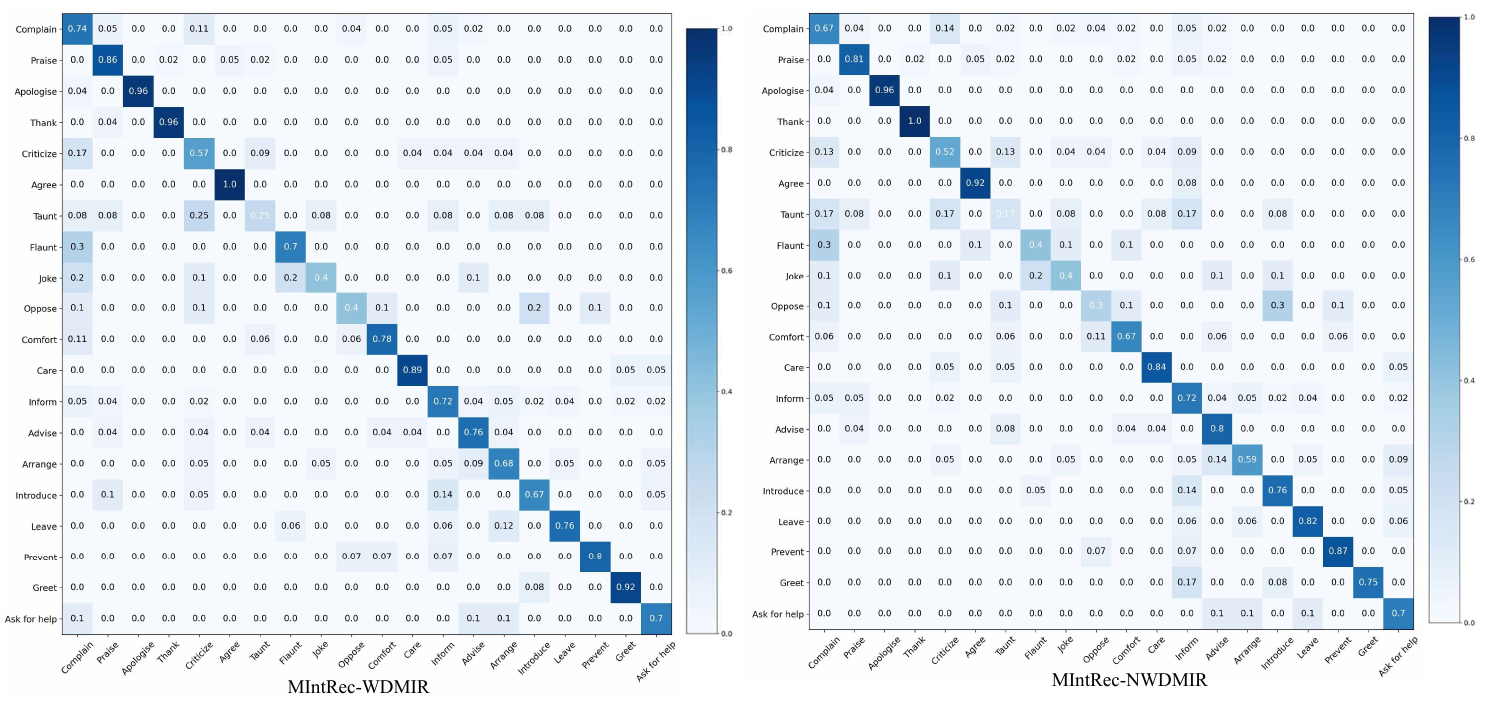}
        \caption{MIntRec-WDMIR is the confusion matrix with wavelet-driven fusion module on MIntRec; MIntRec-NWDMIR is the confusion matrix with wavelet-driven fusion module removed on MIntRec}
        \label{figure 3}
    \end{minipage}
    \hfill
    \begin{minipage}[t]{0.48\textwidth}
        \centering
        \includegraphics[width=\linewidth]{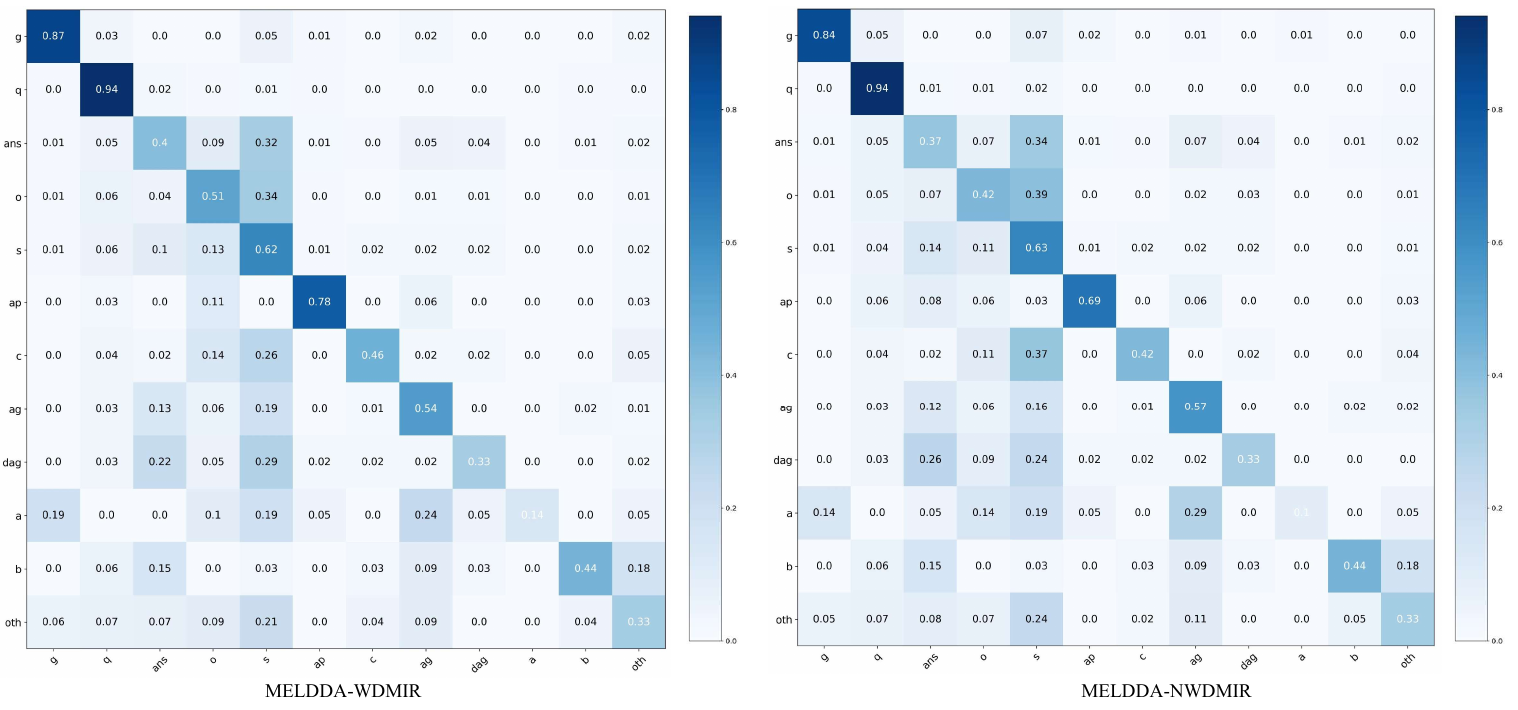}
        \caption{MELDDA-WDMIR is the confusion matrix with wavelet-driven fusion module on MELD-DA; MELDDA-NWDMIR is the confusion matrix with wavelet-driven fusion module removed on MELD-DA}
        \label{figure 4}
    \end{minipage}
\end{figure*}
\begin{table*}
    \centering
    \renewcommand{\arraystretch}{1.1}
    \begin{adjustbox}{max width=\textwidth}
    \begin{tabularx}{\textwidth}{*{14}{>{\centering\arraybackslash}X}}
        \toprule
        \multicolumn{6}{c}{Modules}  & \multicolumn{4}{c}{MIntRec} & \multicolumn{4}{c}{MELD-DA} \\
        \cmidrule(lr){1-6} \cmidrule(lr){7-10} \cmidrule(lr){11-14}
         $F_{lv}$ &$F_{la}$ &$F^{'}_{vat}$ &$F^{'}_{avt}$ &$F_{tva}$ &WFM &ACC &WF1 &WP &R  &ACC &WF1 &WP &R\\
        \midrule
        \texttimes & \texttimes & \checkmark & \checkmark & \checkmark & \checkmark & 73.49 & 73.24 & 73.46 & 70.68 & 61.68 & 60.56 & 60.65 & 51.20 \\
        \checkmark & \checkmark & \texttimes & \texttimes & \checkmark & \checkmark & 72.70 & 72.29 & 72.36 & 69.67 & 60.61 & 59.15 & 59.38 & 49.59 \\
        \checkmark & \checkmark & \checkmark & \checkmark & \texttimes & \checkmark & 72.02 & 71.71 & 71.82 & 69.06 & 61.06 & 59.63 & 59.66 & 50.61 \\
        \checkmark & \checkmark & \checkmark & \checkmark & \checkmark & \texttimes & 72.02 & 71.76 & 72.20 & 68.77 & 60.56 & 59.28 & 59.13 & 49.68 \\
        \midrule
         \checkmark & \checkmark & \checkmark & \checkmark & \checkmark & \checkmark & \textbf{75.05} & \textbf{74.96} & \textbf{75.26}  & \textbf{72.65} & \textbf{62.16} & \textbf{60.94} & \textbf{61.44} & \textbf{51.36} \\
        \bottomrule
    \end{tabularx}
    \end{adjustbox}
     \caption{Conducting ablation studies on the MIntRec and MELD-DA datasets respectively. Features obtained from the first layer are denoted as $F_{lv}$ and $F_{la}$, features from the second layer are denoted as $F^{'}_{vat}$ and $F^{'}_{avt}$, and features from the third layer are denoted as $F_{tva}$. WFM stands for the audio-video feature fusion module.}
     \label{table 2}
\end{table*}
\subsection{Main Result} Our method is compared with the optimal method in \S \ref{Baselines}, and the experimental results are shown in Table \ref{table 1}. From the analysis of the experimental results, it is clear that our approach has made significant progress in two main areas.
First, on the MIntRec dataset, our method achieves 1.13\%, 1.07\%, 1.08\%, and 0.99\% improvement in four key performance metrics, namely, ACC, WF1, WP, and R, respectively, when compared to the best existing baseline method. These results strongly demonstrate the effectiveness of the method in handling multimodal intent recognition tasks in complex real-world scenarios.
Second, on the multi-round sentiment dialog analysis dataset MELD-DA, our method also demonstrates better performance, improving 0.41\%, 1.17\%, 1.11\%, and 1.22\% in the four evaluation metrics of ACC, WF1, WP, and R, respectively, compared to the optimal baseline. This result confirms the effectiveness and robustness of our proposed method in the task of conversational emotion recognition.
Overall, the experimental results fully demonstrate that our method outperforms the existing state-of-the-art baseline methods on all evaluation metrics for two representative datasets, which in turn validates the effectiveness and generalizability of the method.
\begin{figure}[ht]
  \centering
  \includegraphics[width=\linewidth]{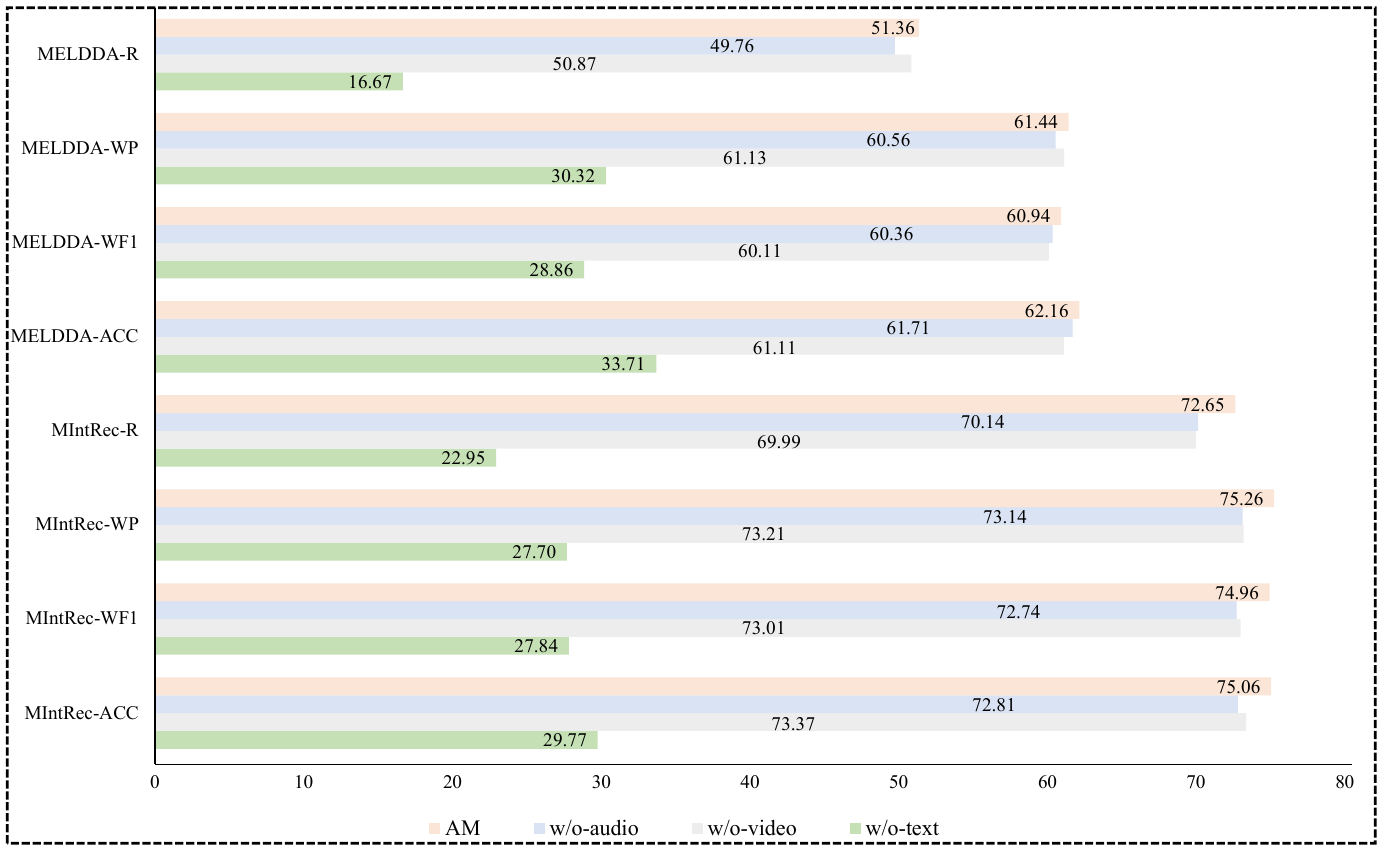}
  \caption{Single-Modality Missing Comparison. AM represents no missing modality, w/o-text indicates the removal of the text modality, w/o-video indicates the removal of the video modality, and w/o-audio indicates the removal of the audio modality.}
  \label{figure 5}
\end{figure}
\begin{figure}[ht]
  \centering
  \includegraphics[width=0.8\linewidth]{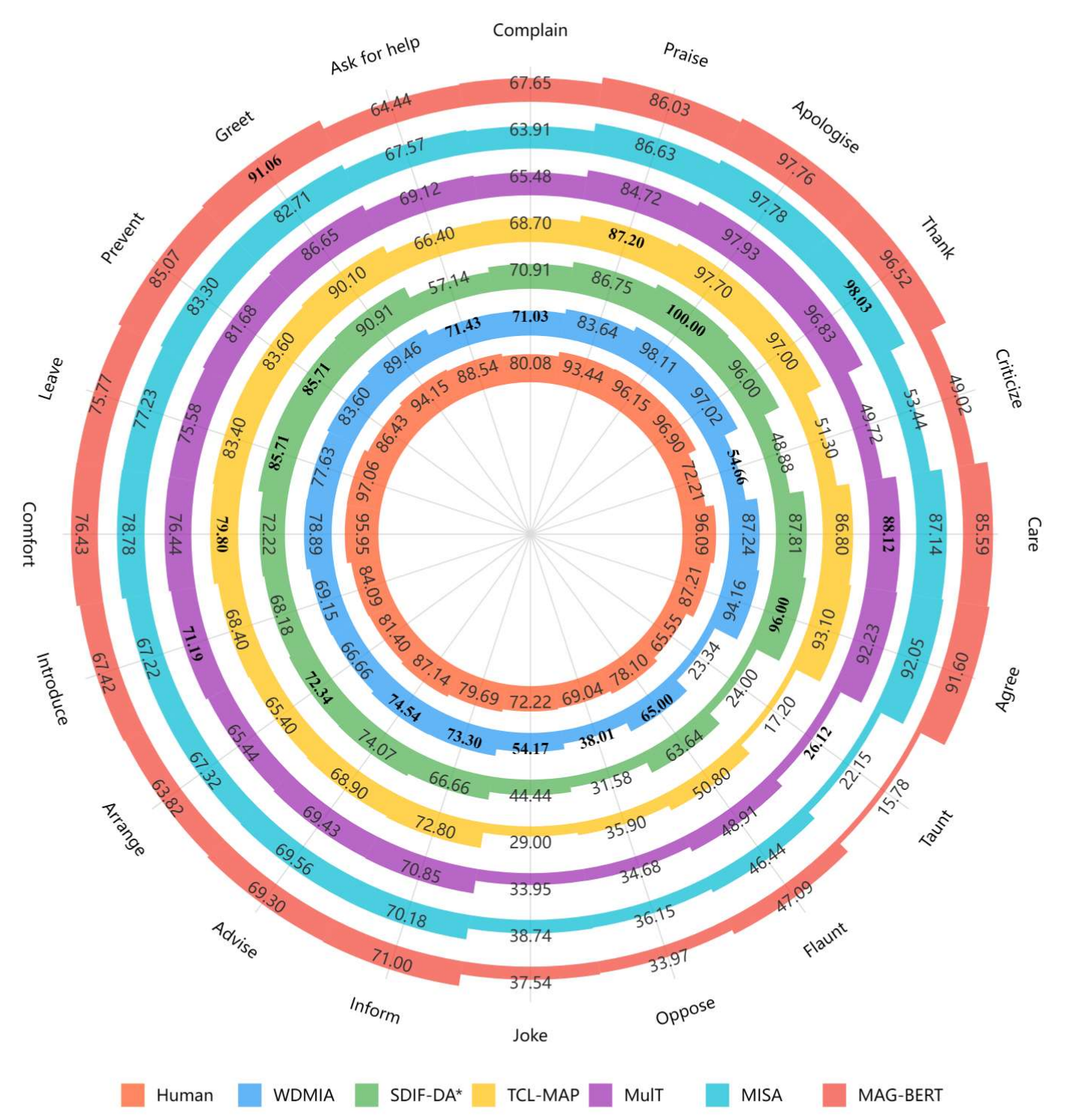}
  \caption{F1-score comparison on the MIntRec dataset between our method and the baseline. SDIF-DA* denotes re-implemented results based on the original paper; other method and human results are from TCL-MAP. Bold indicates the best performance, excluding Human.}
  \vspace{-0.25cm}
  \label{figure 6}
\end{figure}
\subsection{Ablation Study}
To further explore the impact of different modules on the performance of the WDMIR method on the MIntRec and MELD-DA datasets, we perform ablation experiments on the Wavelet-driven Fusion Module, Collaborative Representation Module, and Progressive Fusion Module to determine the contribution of each module to the overall performance.
\subsubsection{Wavelet-driven Fusion Module}
To assess the performance of wavelet-driven fusion module, we conducted an experiment where we excluded it, opting to directly pass the pre-trained video and audio features as inputs to the cross-modal attention mechanism without fusing them. The experimental results are shown in Table \ref{table 2}. The results on the MIntRec dataset showed a decrease in the model's ACC by 3.03\%, WF1 by 3.2\%, WP by 3.06\%, and R by 3.88\%. Similarly, on the MELD-DA dataset, ACC fell by 1.6\%, WF1 by 1.66\%, WP by 2.31\%, and R by 1.68\%. These findings indicate that the wavelet-driven fusion module plays a crucial role in effectively combining video and audio data, thereby enhancing the model's ability to identify positive samples. We can better understand the performance of the wavelet-driven fusion module across different categories on the MIntRec and MELD-DA datasets through the confusion matrices in Figures \ref{figure 3} and \ref{figure 4}. The experimental results indicate that the wavelet-driven fusion module can extract fine-grained semantic information from non-verbal modalities, which has a certain impact on categories with implicit intentions and emotions.
\subsubsection{Collaborative Representation Module}
We explored the impact of the collaborative representation module on WDMIR by removing $F_{tva}$, and the results are shown in Table \ref{table 2}. The experimental results show that ACC is reduced by 3.03\%, WF1 by 3.25\%, WP by 3.44\% and R by 3.59\% on MIntRec. On the MELD-DA dataset, ACC decreased by 1.1\%, WF1 decreased by 1.31\%, WP decreased by 1.78\%, and R decreased by 0.75\%. The collaborative representation module can provide richer and more comprehensive information for intent recognition, capture the complementary information contained in different modalities, and improve the accuracy of intent recognition.
\subsubsection{Progressive Fusion Module}
To verify the impact of the progressive module on the overall performance of WDMIR, we sequentially removed $F_{lv}$ and $F_{la}$, as well as $F^{'}_{vat}$ and $F^{'}_{avt}$. The experimental results are shown in Table \ref{table 2}. The results demonstrate that the progressive fusion module improves model performance on both the MIntRec and MELD-DA datasets. Specifically, $F_{lv}$ and $F_{la}$ compensate for the loss of audio-visual information, while $F^{'}_{vat}$ and $F^{'}_{avt}$ enhance deep fusion between modalities. The ablation experiments further validate the critical role of the progressive fusion module in optimizing multimodal information complementarity and improving the overall performance of the model.
\subsection{Performance with Single-Modality Missing}
In addition to conducting ablation experiments, we also analyzed the impact of single-modality missing scenarios. In Figure \ref{figure 5}, AM represents the case with no missing modality, w/o-text indicates the removal of modules related to the text modality, w/o-video indicates the removal of modules related to the video modality, and w/o-audio indicates the removal of modules related to the audio modality. The vertical axis shows the performance of MIntRec and MELD-DA in terms of ACC, WF1, WP, and R. From the experimental results, it can be observed that removing the text modality in our method leads to a sharp performance drop, indicating that the text modality contains the primary information for intent recognition. Removing the video or audio modalities also results in performance degradation; however, the decline is less severe compared to the text modality, suggesting that the video and audio modalities provide auxiliary information for intent recognition in multimodal tasks.
\subsection{F1-Score Analysis Across Intent Categories}
To better evaluate the performance of WDMIR on MIntRec, we compare its f1-score with baselines to assess how each method performs across fine-grained intent categories. As illustrated in Figure \ref{figure 6}, our approach demonstrates strong performance, particularly in the intent categories of ``Flaunt", ``Joke" and ``Oppose" suggesting that WDMIR is adept at managing complex intent categories. WDMIR performs similarly to other methods in the ``Care" and ``Complain" categories, and even outperforms in some areas, indicating stable performance in common intent categories, likely due to the wavelet-driven nonverbal modal fusion. However, despite some improvements in certain intent categories, WDMIR's performance in categories like ``Taunt", ``Oppose" and ``Joke" is not as strong as that of other methods, which may be linked to the influence of wavelet-driven nonverbal modal fusion. The F1-scores metric performed significantly below human levels on the ``Taunt", ``Oppose", and ``Joke" intentions. This suggests that WDMIR still faces challenges in understanding emotions that are heavily dependent on context and humor, and that further improvements in understanding linguistic humor and contextual variation are needed.
\section{Conclusion}
In this paper, wavelet transform is introduced into the field of multimodal intent recognition for the first time, and a novel wavelet-based multimodal intent recognition method, WDMIR, is proposed. The method serializes multilevel decomposition of video and audio data by wavelet transform and realizes the extraction and fusion of video and audio features in the frequency domain. Through the collaborative fusion module, it realizes the collaborative representation from bimodal to trimodal, which improves the expression of deep features and better identifies the semantic information in multimodal data. Through the progressive fusion module, the whole different levels of feature representations are effectively represented to improve the accuracy of multimodal intent recognition. Comprehensive experimental results show that the proposed wavelet-driven approach can enhance the performance of the model in multimodal intention recognition tasks. This study not only demonstrates the potential of wavelets in multimodal learning but also provides a new perspective on feature extraction and fusion of multimodal data.
\section*{Acknowledgments}
This work was supported by the National Natural Science Foundation of China (Grants No. 62337001, U23A20319, 62406303), the Key Technologies R\&D Program of Anhui Province (No. 202423k09020039), the Anhui Provincial Natural Science Foundation (No. 2308085QF229), and the Fundamental Research Funds for the Central Universities. Additional support was provided by the Youth Science and Technology Talent Growth Project of Guizhou Provincial Department of Education (Grant No. KY[2022]054), the Guizhou Provincial Higher Education Undergraduate Teaching Content and Curriculum System Reform Project (Grant No. GZJG2024331), and the Guizhou Provincial Science and Technology Projects (Grant No. QKHJC[2024]Youth012).

\bibliographystyle{named}
\bibliography{ijcai25}
\end{document}